\documentclass[12pt,oneside,openany]{article}
\usepackage{amsfonts,amssymb,graphicx}
\usepackage{cite}
\usepackage[affil-it]{authblk}
\usepackage{indentfirst}
\usepackage{newlfont}
\usepackage{amsthm}
\usepackage{amsmath}
\usepackage{latexsym}
\usepackage{eucal}
\usepackage{amssymb}
\usepackage{mathrsfs}
\usepackage{graphicx}

\newcommand{\be}{\begin{equation}}
\newcommand{\ee}{\end{equation}}

\begin{document}
\title{Inverse problem for multi-species mean field models in the low temperature phase}

\author[1]{Micaela Fedele}
\affil[1]{Dipartimento di Matematica, Universit\`a di Bologna}
\author[2]{Cecilia Vernia}
\affil[2]{Dipartimento di Scienze Fisiche Informatiche e Matematiche, Universit\`a di Modena e Reggio Emilia}

\date{}

\maketitle

\begin{abstract}
In this paper we solve the inverse problem for a class of mean field models (Curie-Weiss model and its multi-species version) when multiple thermodynamic states are present, as in the low temperature phase where the phase space is clustered. The inverse problem consists in reconstructing the model parameters starting from configuration data generated according to the distribution of the model. We show that the application of the inversion procedure without taking into account the presence of many states produces very poor inference results. This problem is overcomed using the clustering algorithm. When the system has two symmetric states of positive and negative magnetization, the parameter reconstruction can be also obtained with smaller computational effort simply by flipping the sign of the magnetizations from positive to negative (or viceversa).  
The parameter reconstruction fails when the system is critical: in this case we give the correct inversion formulas for the Curie-Weiss model and we show that they can be used to measuring how much the system is close to criticality.
\end{abstract}

\section{Introduction}
In the statistical physics literature of the last decades a growing attention has been devoted to the study of the inverse problem\cite{AuEk,Ngu,CaBi,SM}. This amounts to study how to infer the parameters of a model starting from the observation of real data. In particular, the application of the inverse Ising model, although known for a long time as Boltzmann machine learning\cite{ahs,T}, has aroused interest in recent years in many different  fields (physics\cite{AuEk,Ngu}, neuroscience\cite{CoMo,RTH}, biology\cite{WWSHH,BCGMSVW}, social and health sciences\cite{CG,BCSV,ABCSV,BCFVV}), especially since the advent of the big-data age.  In these applications, stemming from the assumption that  the real world system of interest is described by an Ising model with hamiltonian $H$,  the inverse problem amounts to fit $H$ to the system, i.e. to  calculate the parameters of the underlying $H$ from experimentally measured expectation values.

In this paper we consider the inverse problem for the Curie-Weiss model and for its multi-species version\cite{GC}. These models, among all the possible choices, have the advantage of being very simple and thus of allowing for analytical computations, but still sufficiently general to represent a wide range of interesting phenomena. In fact, recent studies has shown that such models provide surprisingly accurate descriptions of real world phenomena\cite{BCFVV}. The Curie-Weiss hamiltonian depends on the coupling parameter and the external magnetic field that can be efficiently inferred, in the uniqueness region of the model, from the estimates of the magnetization and the susceptibility obtained by a sample of spin configurations, as shown in Ref.~\cite{FVC}. Here we take a step forward by considering how to solve the inverse problem when the consistency equation has more than one solution. The presence of many states in the phase space can occur, for example, when the system undergoes a phase transition. In this case, the clustered structure of the sampled input configurations may produce bad coupling parameter inference.
In fact, in ferromagnetic systems below the ferromagnetic transition the configurations are grouped in two clusters of positive and negative magnetization. 
We show that coupling parameters can be well
inferred also in the low temperature phase in two ways: either globally by applying the inverse problem procedure to the whole set of the input configurations after changing the sign of the magnetizations from positive to negative (or viceversa) or locally, by clustering the configurations and then applying the algorithm separately to data in each cluster. 
While this last method, known in literature as clustering algorithm\cite{Mac,RoLa,DeTe}, is general and can be used with different models that exhibits multiple states, the sign flip is suitable only for models with couples of symmetric solutions. 
In a recent study\cite{CLV}, the clustering algorithm has been used to solve the inverse problem for the model of interacting monomer-dimers on the complete graph, whose solutions are not symmetric in the coexistence phase. The parameter estimates are very accurate and in good agreement in both ways, although the clustering algorithm has higher computational cost.

Following the methods used in Ref.~\cite{FVC}, we validate the inversion procedure that we propose here, by sampling a set of spin configurations from the equilibrium distribution of the model, and by reconstructing the underlying parameters from a large number of such samples. 
When dealing with real phenomenological data the solution of the inverse problem requires first to provide the explicit expression of the model free parameters with respect to the macroscopic thermodynamic variables and then to evaluate these macroscopic variables starting from the the data. The first step is obtained considering the consistence equation of the model\cite{FVC}, the second one with the maximum likelihood estimation procedure\cite{Fi,J,J2}.

Finally, we show that if the system is critical, the analytical inversion formulas do not apply and the parameter estimation fails.

\section{Inverse Problem for the Curie-Weiss Model}\label{CWSec}
The Curie-Weiss model for a system of $N$ spin particles is defined by the Hamiltonian:
\begin{equation}\label{Hami.curie}
H_{N}(\sigma)=-\frac{J}{2N}\sum_{i,j=1}^{N}\sigma_{i}\sigma_{j}-h\sum_{i=1}^{N}\sigma_{i}
\end{equation}
where $\sigma_{i}\in\{+1,-1\}$ is the spin of the $i$-th particle, $J>0$ is the coupling constant and $h$ is the magnetic field.
The probability of a configuration of spins $\sigma=(\sigma_{1},\dots,\sigma_{N})$ is given by the Boltzmann-Gibbs measure:
\begin{equation}\label{BG.curie}
P_{N, J, h}\{\sigma\}=\dfrac{e^{-\beta H_{N}(\sigma)}}{\sum_{\sigma\in\{\pm 1\}^{N}}e^{-\beta H_{N}(\sigma)}}  
\end{equation}
where $\beta$ is the inverse temperature. The main observable of the model is the total magnetization, obtained by computing the arithmetic mean of the spins:
\begin{equation}
m_{N}(\sigma)=\frac{1}{N}\sum_{i=1}^{N}\sigma_{i}.
\end{equation}
The behavior of $m_{N}(\sigma)$ in the limit of an infinite number of particles is fully described by the stable solutions of the consistence equation\cite{E}:
\begin{equation}\label{eqcons}
m=\tanh(\beta(Jm+h)). 
\end{equation}
In particular, the average value of $m_{N}(\sigma)$ with respect to the Boltzmann-Gibbs measure, $\langle m_{N}(\sigma) \rangle$, is equal, in the thermodynamical limit, to the mean of such stable solutions. When the magnetic field is absent the number of stable solutions depends on the product between the coupling constant and the inverse temperature. For $\beta J\leq 1$ the consistence equation admits a unique solution, stable, in the origin; for $\beta J>1$ the origin becomes unstable while other two stable solutions $\pm m^{*}$ arise. In both cases $\langle m_{N}(\sigma) \rangle$ is equal to zero in the limit $N\to\infty$. When the field is different from zero the consistence equation admits always a unique stable solution with the same sign of the field. Such a solution is not always the only possible one; in fact, Eq.~(\ref{eqcons}) allows also the presence of a metastable solution and of an unstable solution. With the exception of the case of $\beta J=1$ and $h=0$, we can write the model parameters as follows:
\begin{align}
J&=\frac{1}{\beta(1-m^{2})}-\frac{1}{\chi} \label{JTerm} \\ 
h &=\frac{\tanh^{-1}(m)}{\beta}-Jm \label{hTerm} 
\end{align}
where $m$ is a stable solution of (\ref{eqcons}) and $\chi=\partial m/\partial h$ is the susceptibility of the system. When $\beta J=1$ and $h=0$, Eq.~(\ref{JTerm}) and (\ref{hTerm}) become meaningless because the susceptibility grows to infinity. This critical case is analyzed in detail in section \ref{CWcriticoSec}. In the following, for the sake of simplicity, we consider the inversion temperature $\beta$ absorbed within the model parameters. This is analogous to fix its value equal to $1$.

We mentioned above that as $m$ is a unique stable solution of the consistence equation, $\langle m_{N}(\sigma) \rangle$ tends to such a value as $N$ grows to infinite. In this case, $\chi$ represents the infinite volume limit of the product between the variance of the total magnetization, $\langle m_N^2(\sigma)\rangle - \langle m_N(\sigma)\rangle^2$, and the number of spins $N$. Therefore, by estimating these macroscopic quantities from the data and using identities (\ref{JTerm}) and (\ref{hTerm}), we can infer the values of the model parameters. In the following, we call finite size magnetization 
\begin{equation}\label{bram}
m_N=\langle m_{N}(\sigma) \rangle
\end{equation} 
 and finite size susceptibility 
\begin{equation}
\chi_N = N\left(\langle m_N^2(\sigma)\rangle - \langle m_N(\sigma)\rangle^2\right).
\end{equation} 
When there are two stable solutions of (\ref{eqcons}), $m_N$ is equal to zero by symmetry. As a consequence, its estimation from the data does not allow us to compute the true model parameters. In section \ref{2stableSol} we show how it is possible to solve the inverse problem also in this case. 

In the case of a unique stable solution of (\ref{eqcons}), in order to estimate the parameters, we need a sample of $M$ independent spin configurations, $\sigma^{(1)}, \sigma^{(2)}, \dots,  \sigma^{(M)}$, distributed according to (\ref{BG.curie}). Starting from the total magnetization
\begin{equation}
m_{N}(\sigma^{(s)})=\frac{1}{N}\sum_{i=1}^{N}\sigma_{i}^{(s)} 
\end{equation}
of each spin configuration, we use the maximum likelihood procedure to compute the estimators of $m_N$ and $\chi_N$, as follows:
\begin{equation}\label{mexpchiexp}
m_{exp}=\dfrac{1}{M}\sum\limits_{s=1}^{M}m_{N}(\sigma^{(s)}), \quad\quad \chi_{exp}=N\left(\dfrac{1}{M}\sum\limits_{s=1}^{M}m_{N}^{2}(\sigma^{(s)})-m_{exp}^{2}\right).
\end{equation}
This method determines the free parameters of the distribution, by imposing that their values maximize the probability to obtain the given sample of spin configurations.
Eventually, by combining (\ref{mexpchiexp}) with (\ref{JTerm}) and (\ref{hTerm}) we obtain the free parameter estimators:
\begin{align}
J_{exp} &=\frac{1}{1-m_{exp}^{2}}-\frac{1}{\chi_{exp}} \label{stimatori}\\
h_{exp} &=\tanh^{-1}(m_{exp})-J_{exp}m_{exp}.\label{stimatorih}
\end{align}

As a general remark, note that the parameter estimation involves two kinds of approximations: one in the inverse problem formulas (\ref{JTerm}) and (\ref{hTerm}), that require $m$ and $\chi$, i.e. the infinite volume limit of $m_N$ and $\chi_N$, the other in the statistical evaluation of $m_N$ and $\chi_N$ through $m_{exp}$ and $\chi_{exp}$ with the maximum likelihood estimation procedure given in (\ref{mexpchiexp}). The accuracy of the first approximation increases with $N$, that of the second one with $M$. The evidence of these two facts together with the numerical thresholds for the choices of $N$ and $M$ for the Curie-Weiss model were deeply investigated in Ref.~\cite{FVC} with some numerical tests. 

When the solution of (\ref{eqcons}) is no more unique, the inversion procedure presented above is no longer suitable, as it will be clear in what follows. Therefore, we need to consider alternative algorithms to address and solve the problem.
In the next sections, we present numerical tests in order to validate the inversion procedure both for the case when the phase space presents only one state and when the system undergoes a phase transition.

\subsection{Numerical tests}
The aim of this work is to show the robustness of the inverse problem for experiments with real world datasets; thus we fix $M=1000$ and consider $N\in [100, 10000]$. This choice for the sizes of the sample $M$ and of the system $N$ is an acceptable compromise between the requirement of stabilizing the estimators and the simulation of a realistic experimental dataset.

From the numerical point of view, fixed the values of the system size $N$ and of the parameters $J$ and $h$, we extract each configuration from a virtually exact simulation of the equilibrium distribution (\ref{BG.curie}). In fact, due to the mean field nature of the model, the Boltzmann-Gibbs distribution of the total magnetization $m_N(\sigma)$ can be computed by evaluating the combinatorial weights $C_m$ of its possible values $m\in\{-1, -1+\frac{2}{N}, -1+\frac{4}{N}, \dots, 1-\frac{2}{N}, 1\}$ as: 
\begin{equation}
P_{N, J, h}\{m_N(\sigma)=m\}=\dfrac{C_m\exp(\frac{J}{2}m^2+hm)}{\sum_{m}C_m\exp(\frac{J}{2}m^2+hm)}  
\end{equation}
where 
\begin{equation}
C_m = \dfrac{N!}{\left(\frac{N(1+m)}{2}\right)!\left(\frac{N(1-m)}{2}\right)!}
\end{equation}
gives the number of spin configurations that share the same value $m$ of the total magnetization. We use the probability distribution obtained in this way to extract large samples of magnetizations that will be used in (\ref{mexpchiexp}) to compute $m_{exp}$ and $\chi_{exp}$.

Moreover, in order to assess the statistical dependence of the estimators on the sample
$(\sigma^{(1)}, \sigma^{(2)}, \dots,  \sigma^{(M)})$ we consider $20$ independent instances of such a sample, we apply the maximum likelihood estimation to each of them independently and then we average over the $20$ inferred values. In what follows we use the subscript $exp$ to denote the estimators (i.e. $m_{exp}$, $\chi_{exp}$, $J_{exp}$ and $h_{exp}$) and the bar symbol ($\bar{m}_{exp}$, $\bar{\chi}_{exp}$, $\bar{J}_{exp}$ and $\bar{h}_{exp}$) for their statistical mean over the $20$ $M$-samples. We find numerical evidence\cite{FVC} that it suffices $M\ge 1000$ in order to obtain acceptable parameter estimations.

%

Taking into account the description of the number of solutions of Eq.~(\ref{eqcons}) given in section \ref{CWSec}, in what follows we test numerically the inverse problem for all the different possible cases.

\subsubsection{Case of a unique solution of the consistence equation}
\begin{figure}[h!]
\centering
\includegraphics[width=0.7\textwidth]{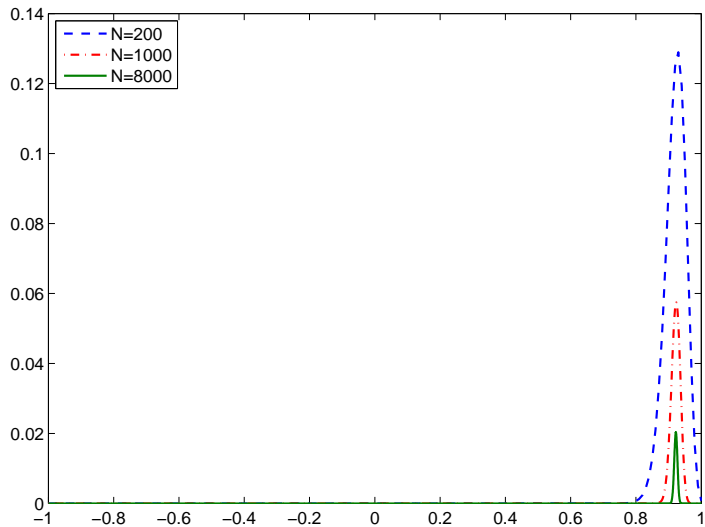}
\caption{\label{Fig1} Boltzmann-Gibbs distribution of the total magnetization $m_N(\sigma)$ for $J=1.5$, $h=0.22$ and different values of the number of spins $N$. The distribution is given by the blue dashed line for $N=200$, by the red dot-dashed line for $N=1000$ and by the green continuous line for $N=8000$. The peak of the distribution is centered around the solution $m^*=0.922$ of the consistency equation (\ref{eqcons}).}
\end{figure}

Let us start considering a couple of parameters $(J,h)$ for which there is only one solution $m^*\in [-1, 1]$ 
of Eq.~(\ref{eqcons}). In this case the Boltzmann-Gibbs distribution of the total magnetization $m_N(\sigma)$ presents a unique peak centered around the solution $m^*$, as shown in Fig.~\ref{Fig1} for the case $J=1.5$ and $h=0.22$, where $m^*=0.922$. As $N$ increases, the peak shrinks towards the value of the solution, meaning that its estimation through the finite size magnetization $m_N$ becomes more and more accurate.

\begin{figure}[h!]
\centering
\includegraphics[width=0.7\textwidth]{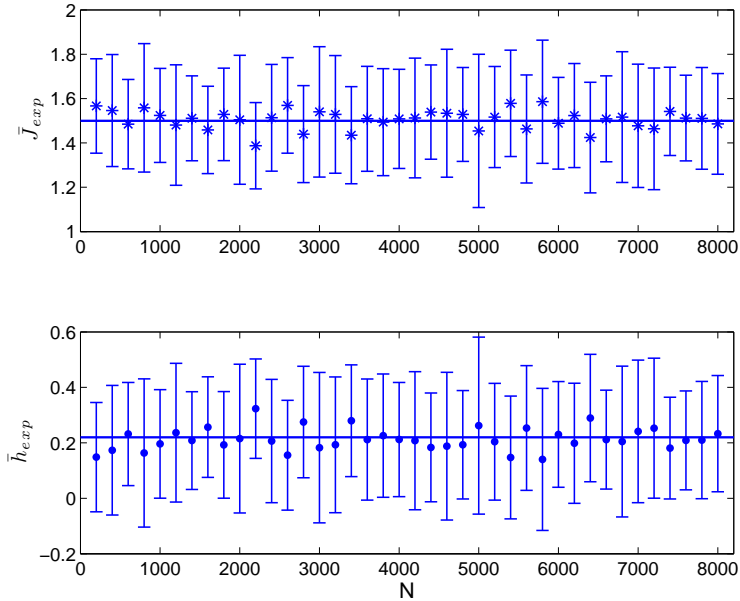}
\caption{\label{Fig2} $\bar{J}_{exp}$ (upper panel) and $\bar{h}_{exp}$ (lower panel) as a function of $N$ for $J=1.5$, $h=0.22$ and $M=1000$. Error bars are standard deviations on $20$ different $M$-samples of configurations of the same system (see text for the details of the simulation). The horizontal lines correspond to the exact values of $J=1.5$ (upper panel) and $h=0.22$ (lower panel).}
\end{figure}

For this case, the estimations of  
$J_{exp}$ and $h_{exp}$ are plotted in Fig.~\ref{Fig2} as functions of $N\in[200, 8000]$. Note that the inferred values of $\bar{J}_{exp}$ and $\bar{h}_{exp}$ are in optimal agreement with the exact values of the parameters (continuous lines in Fig.~\ref{Fig2}), even when the size $N$ of the system is very small. Moreover, the error bars obtained with the standard deviation on the $20$ different $M$-samples of configurations of the same system are comparable for all the considered values of $N$.

\begin{figure}[h!]
\centering
\includegraphics[width=0.7\textwidth]{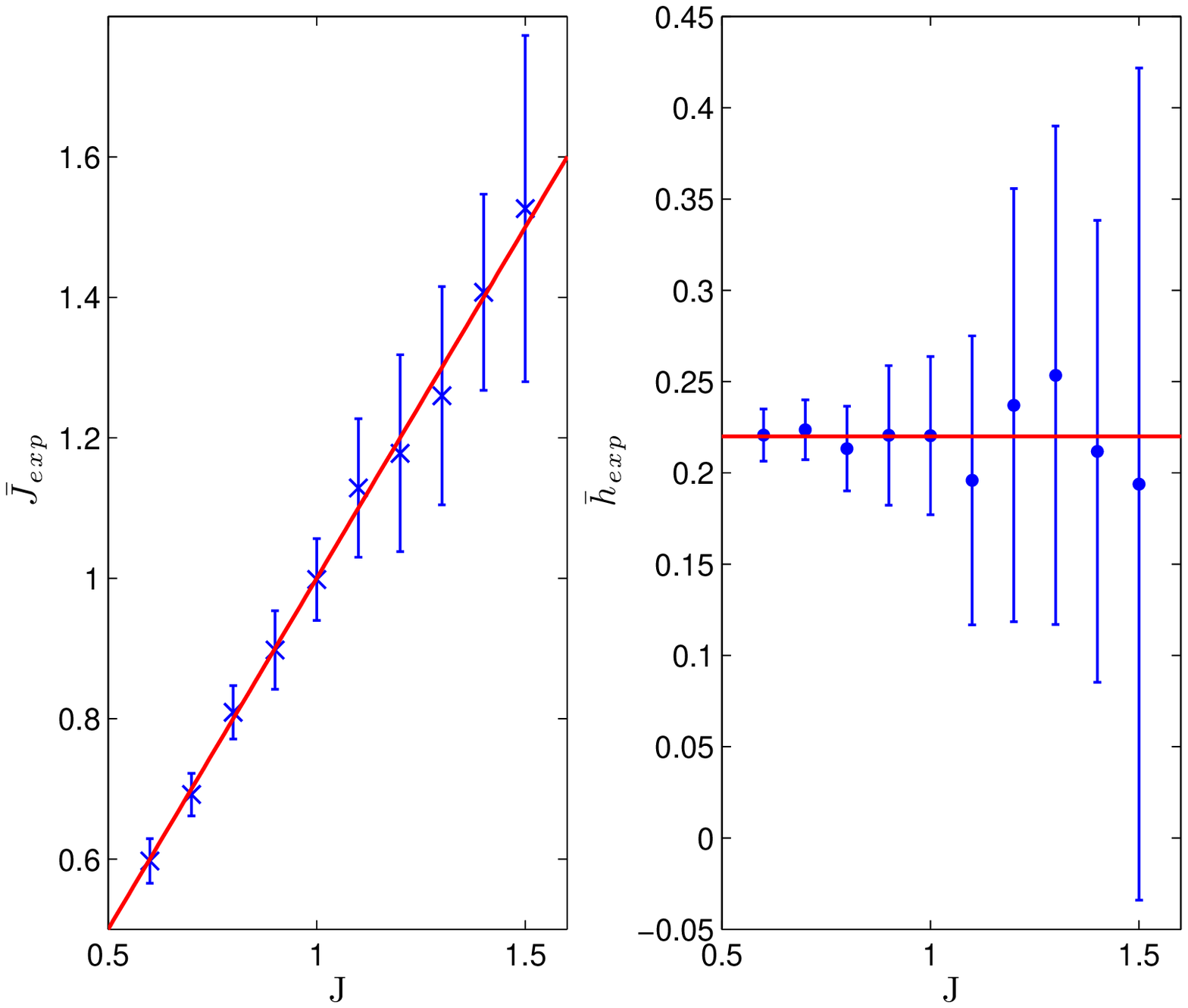}
\caption{\label{FigErrCreJ} $\bar{J}_{exp}$ (left panel) and $\bar{h}_{exp}$ (right panel) as a function of $J$ for $h=0.22$ and $N=1000$. Error bars are standard deviations on $20$ different $M$-samples of configurations of the same system (see text for the details of the simulation). The red continuous line in the left panel represents $\bar{J}_{exp}=J$, the red horizontal line in the right panel corresponds to the exact value of the magnetic field $h=0.22$.}
\end{figure}

Fig.~\ref{FigErrCreJ} shows the parameter estimation as a function of the interacting parameter $J$ for a fixed nonzero value of the magnetic field. Observe that the reconstruction is good also for $J > 1$, but the error bars increase greatly because the interaction between particles is growing.

\subsubsection{Case of two stable solutions of the consistence equation}\label{2stableSol}
\begin{figure}[h!]
\centering
\includegraphics[width=0.7\textwidth]{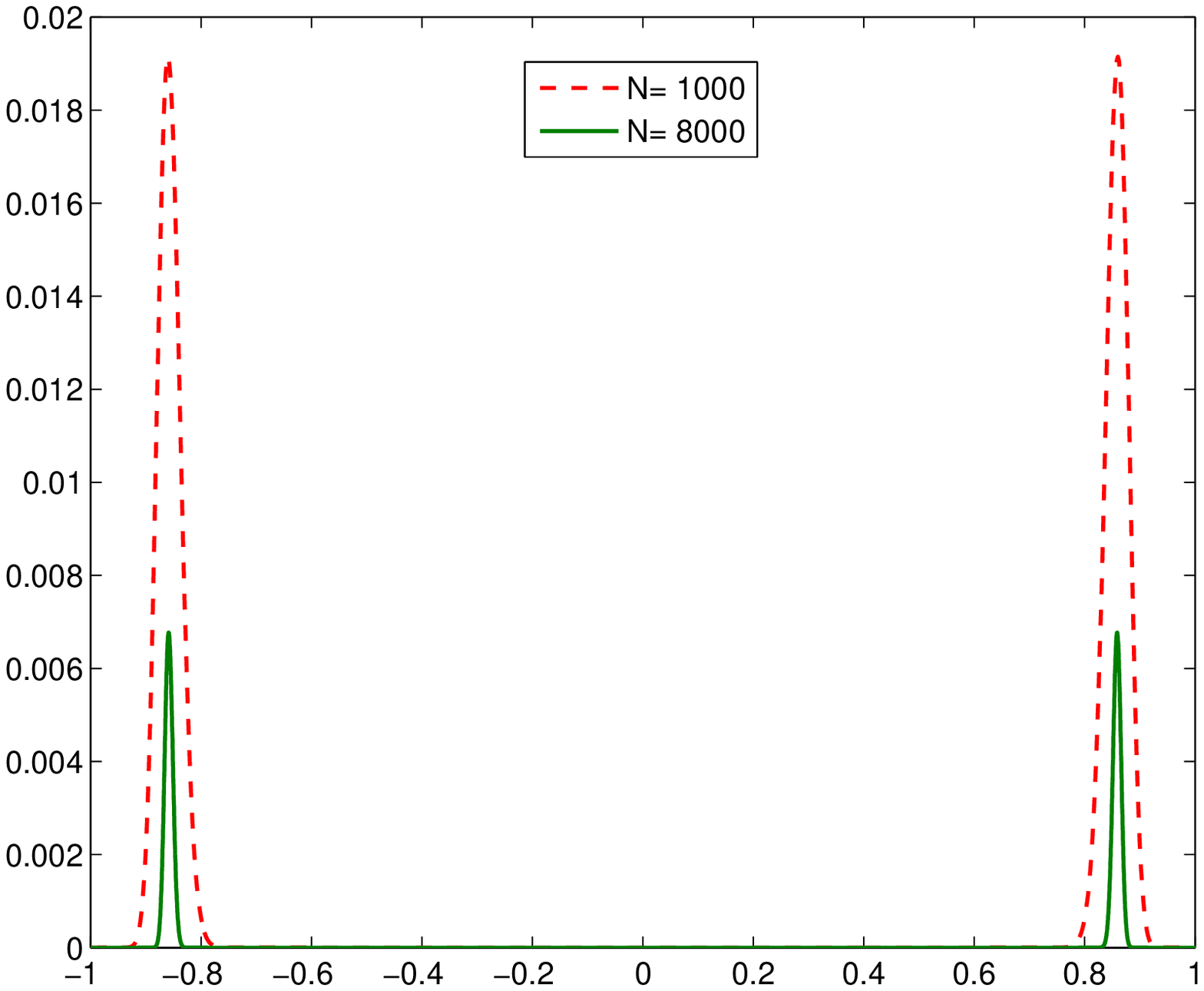}
\caption{\label{Fig4a} Boltzmann-Gibbs distribution of the total magnetization for $J=1.5$, $h=0$ and different values of the number of spins $N$. The distribution is given by the red dot-dashed line for $N=1000$ and by the green continuous line for $N=8000$. The peaks of the distribution are centered around the two symmetric solutions $\pm m^*$  of the consistency equation (\ref{eqcons}), with $m^*=0.8586$.}
\end{figure}

Let us consider now the case in which the consistency equation admits two stable solutions $\pm m^{*}$, that happens as the magnetic field is equal to zero and $J$ is bigger than $1$. In this case the Boltzmann-Gibbs distribution of the total magnetization presents two peaks, both for the finite size systems and in the thermodynamic limit:  one peak is in correspondence with the negative solution $-m^{*}$ of (\ref{eqcons}) and one in correspondence with the positive solution $+m^{*}$ (see Fig.~\ref{Fig4a} as an example for finite size systems). As a consequence, the finite size magnetization $m_N$, as defined in (\ref{bram}), is equal to zero by symmetry and does not tend to one of the two stable solutions of (\ref{eqcons}) when $N$ grows to infinity. 

\begin{figure}[!h]
\centering
\includegraphics[width=0.8\textwidth]{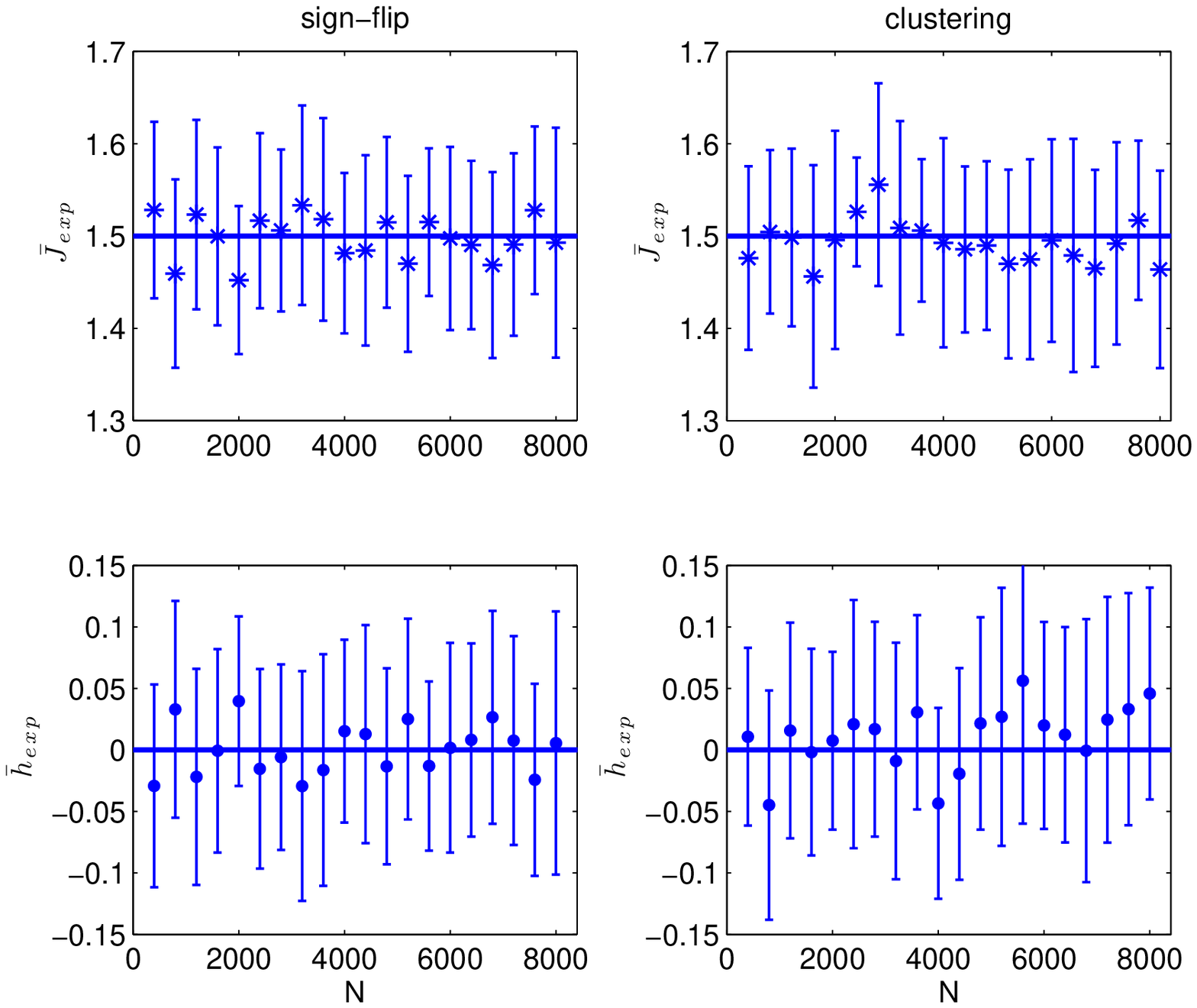}
\caption{\label{Fig13} Values of $\bar{J}_{exp}$ (upper panels) and $\bar{h}_{exp}$ (lower panels) as a function of $N$ for $J=1.5$ and $h=0$ obtained by changing the sign of the negative magnetizations before to apply the inversion procedure (left panels) and with the clustering algorithm (right panels). Error bars are standard deviations on $20$ different independent $M$-samples of the same system ($M=1000$ - see text for the details of the simulation). The horizontal lines correspond to the exact values of $J=1.5$ (upper panels) and $h=0$ (lower panels).}
\end{figure}

Therefore, the inverse problem approach shown in section \ref{CWSec} can not be used to reconstruct the model parameters. Nevertheless, since the inversion formulas \ref{JTerm} and \ref{hTerm} hold true both for $m=-m^{*}$ and for $m=+m^{*}$ we need only to estimate properly at least one of such values from the data. This can be achieved by changing the sign of the negative (positive) experimental magnetizations and then by applying the inversion procedure to the obtained $M$-sample with all positive (negative) magnetizations. The result of the sign-flip is shown in the left panels of Fig.~\ref{Fig13} for $J=1.5$.

The simple trick of inverting the sign of the magnetizations in the sampled input configurations of the inverse problem is possible only if the system has symmetric solutions. 
For this reason, in scientific literature, the procedure used to handle the case of more than one stable solution of Eq.~(\ref{eqcons}), is the clustering algorithm\cite{DeTe}. This procedure has the advantage of being of general application and not only suitable for symmetric models as shown in Ref.~\cite{CLV}. Without going into detail and referring to Refs.~\cite{Ngu,Mac,RoLa} for a depth study on this topic, we only mention that the clustering algorithm divides the configurations in groups (clusters) based on the measure of their mutual distance: configurations belong to the same cluster if their distance is below an appropriate fixed threshold.
In particular, the algorithm defines the density around each configuration as the number of configurations in the given range and put each of them in the cluster with higher density among the closest ones.
This procedure depends obviously on the arbitrary choice of the threshold. 
In our case, the algorithm allows the identification of two clusters $C_1$ and $C_2$. By computing the values of $m_{exp}$ and $\chi_{exp}$ for each of them and then applying (\ref{JTerm}) separately to each cluster, we obtain two different estimators $J_{C_1}$, $J_{C_2}$. Finally, their average weighted with the number $M_k, k=1,2$ of configurations in each cluster ($M_1 + M_2 = M$), gives the estimate of the interaction parameter:

\begin{equation}
J_{exp} = \frac{1}{M}\sum_{k=1}^{2}M_kJ_{C_k}.
\end{equation}

Then, to estimate the magnetic field, we use (\ref{hTerm}) within each cluster, obtaining $h_{C_1}$, $h_{C_2}$ and then we compute their weighted average over the clusters to get the estimator:

\begin{equation}
h_{exp} = \frac{1}{M}\sum_{k=1}^{2}M_kh_{C_k}.
\end{equation}

The results obtained with the clustering algorithm are shown in the right panels of Fig.~\ref{Fig13} for $J=1.5$.
It is interesting to observe that in both cases (right and left panels of Fig.~\ref{Fig13})  the results are qualitatively similar and in good agreement with the exact values of the parameters, though using the clustering there is a higher computational cost than with  the sign-flip. 


\subsubsection{Case of a metastable solution of the consistence equation}
\begin{figure}[h!]
\centering
\includegraphics[width=0.7\textwidth]{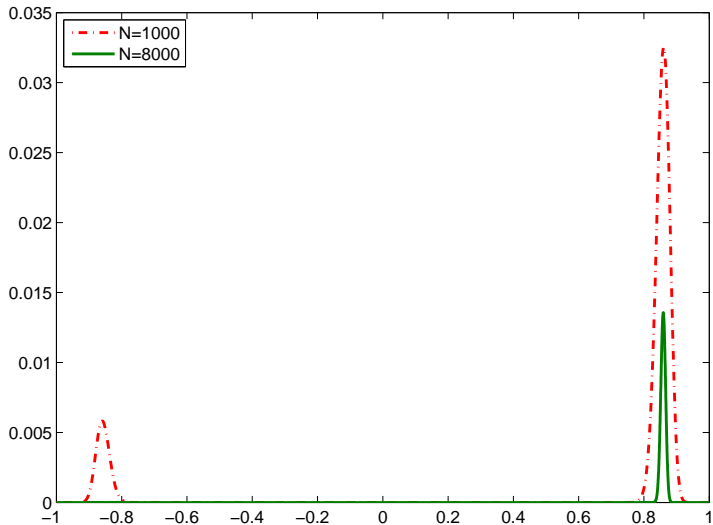}
\caption{\label{Fig4} Boltzmann-Gibbs distribution of the total magnetization for $J=1.5$, $h=0.001$ and different values of the number of spins $N$. The distribution is given by the red dot-dashed line for $N=1000$ and by the green continuous line for $N=8000$. Note that the peaks of the distribution are centered around the two solutions of the consistency equation (\ref{eqcons}), $m_1^*=0.85899$ the stable one and   $m_2^*=-0.85812$ the metastable one,  whose probability vanishes as $N$ goes to infinity (green continuous curve).}
\end{figure}

\begin{figure}[!h]
\centering
\includegraphics[width=0.7\textwidth]{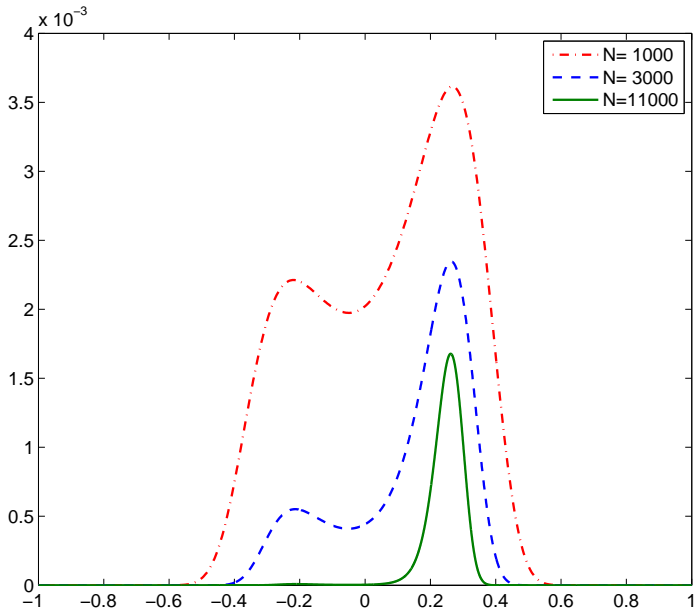}
\caption{\label{Fig14} Boltzmann-Gibbs distribution of the total magnetization for $J=1.02$, $h=0.001$ and different values of the number of spins $N$. The distribution is given by the red dot-dashed line for $N=1000$, by the blue dashed line for $N=3000$ and by the green continuous line for $N=11000$. Note that the peaks of the distribution are centered around the two solutions of the consistency equation (\ref{eqcons}), $m_1^*=0.261727$ the stable one and   $m_2^*=-0.211086$ the metastable one,  whose probability vanishes as $N$ goes to infinity (green continuous curve).}
\end{figure}

Eventually, let us consider the case in which Eq.~(\ref{eqcons}) admits a metastable solution in addition to the stable one, that happens for $J>1$ and $h$ close to zero. In the thermodynamic limit, the Boltzmann-Gibbs distribution of the total magnetization presents a unique peak in correspondence with the stable solution.
However, the presence of the metastable solution in the infinite volume limit is reflected at finite $N$ by the existence of an extra peak in the distribution, as evidenced in Fig.~\ref{Fig4} and in Fig.~\ref{Fig14}, because of the finite size effects. 

\begin{figure}[h!]
\centering
\includegraphics[width=0.7\textwidth]{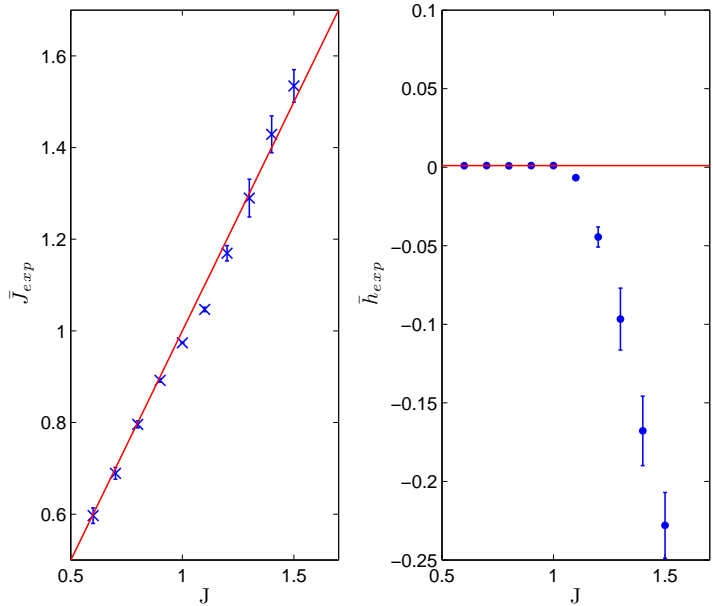}
\caption{\label{FigErrRic} Values of $\bar{J}_{exp}$ (left panel) and $\bar{h}_{exp}$ (right panel) as a function of $J$ for $h=0.001$ and $N=1000$. Error bars are standard deviations on $20$ different $M$-samples of configurations of the same system (see text for the details of the simulation). The red continuous line in the left panel represents $\bar{J}_{exp}=J$, the red horizontal line in the right panel corresponds to the exact value of the magnetic field $h=0.001$.}
\end{figure}

Since for small  $N$ the Boltzmann-Gibbs distribution has a second peak in correspondence with the metastable solution, the application of the standard inversion procedure does not allow the proper reconstruction of the model parameters.
In fact, when $J$ becomes greater than $1$, Fig.~\ref{FigErrRic} shows that the inverse problem formulas lead to very poor results. In particular, note that as $J$ grows from $1$, the values of $\bar{J}_{exp}$ deviate from the exact value of $J$ (red line in Fig.~\ref{FigErrRic}) and that the true magnetic field $h$ is more and more badly estimated. As a last remark, observe that the error bar growth is due to the increase of the interaction (as previously shown in Fig.~\ref{FigErrCreJ} for the case of two stable solutions).

\begin{figure}[h!]
\centering
\includegraphics[width=0.8\textwidth]{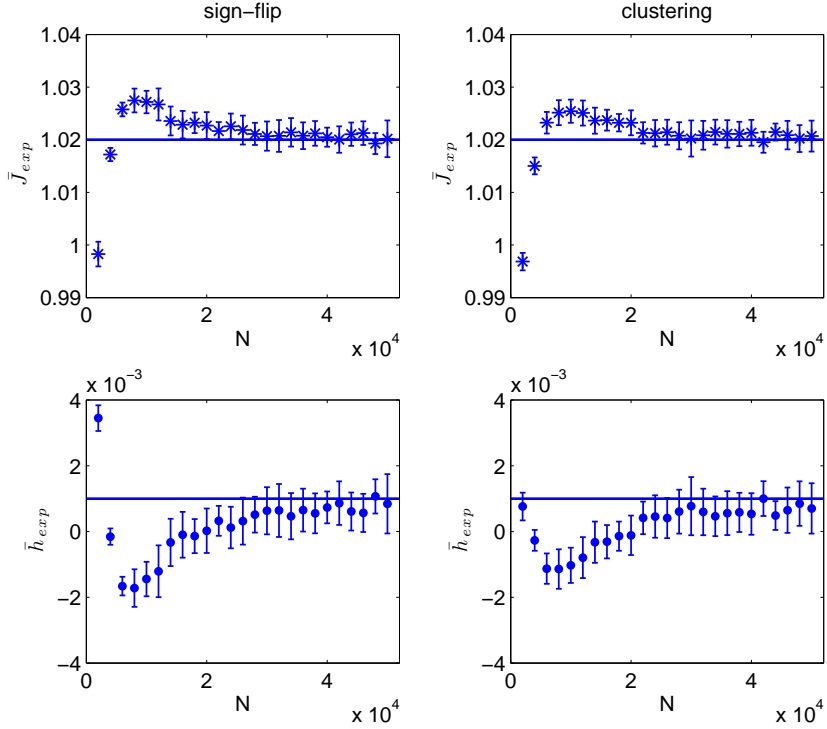}
\caption{\label{Fig16} Values of $\bar{J}_{exp}$ (upper panels) and $\bar{h}_{exp}$ (lower panels) as a function of $N$ for $J=1.02$ and $h=0.001$ obtained by changing the sign of the negative magnetizations before to apply the inversion procedure (left panels) and with clustering algorithm (right panels). Error bars are standard deviations on $20$ different $M$-samples of the same system ($M=1000$ - see text for the details of the simulation). The horizontal lines correspond to the exact values of $J=1.02$ (upper panels) and $h=0.001$ (lower panels).}
\end{figure}

In particular, we can distinguish two different situations depending on the shape of the Boltzmann-Gibbs distribution of $m_N(\sigma)$: a first one in which the supports of the two peaks are disjoint sets (see Fig.~\ref{Fig4}) and a second one in which they are not (see Fig.~\ref{Fig14}). In the first case, the correct estimation of the model parameters is possible by applying one of the two techniques shown in section \ref{2stableSol} for the case of two stable solutions of Eq.~(\ref{eqcons}). In the second one, also the application of such procedures does not allow a proper reconstruction of the parameters, as we can see from Fig.~\ref{Fig16}. Nevertheless, the reconstruction errors both for sign-flip (left panels) and clustering (right panels) are smaller than $2$\% also in the worst case ($N=2000$).
Obviously, when there are stable and metastable solutions with not disjoint supports, the only way to compute efficiently the values of the model parameters is to have a large number $N$ of spins in the sample configuration in order to obtain a better approximation of the thermodynamic limit.

We conclude this section observing that starting with real world experimental dataset, we could be in the case of a metastable solution (or two stable solutions) also when all the magnetizations computed from experimental configurations have the same sign. This could be due to the fact that the data come from a Boltzmann-Gibbs distribution like that of Fig.~\ref{Fig4} (or Fig.~\ref{Fig4a}) conditioned to its positive or negative magnetization peak. In particular, observe that the experimental magnetization can have the same sign of the peak with smaller probability. These are rare events, but still possible if either system size $N$ or the sample size $M$ is too small. 
 In this situation, the parameter estimation is performed with the standard inversion procedure shown in section \ref{CWSec}, but the obtained values are those of a bimodal distribution with one of the two peaks in correspondence to the experimental magnetization $m_{exp}$. Note that if this is the case the sign of the reconstructed value for the magnetic field (when different from zero) could not be in accordance with that of $m_{exp}$ used in the inversion formulas.



\subsubsection{The critical Curie-Weiss model}\label{CWcriticoSec}
\begin{figure}[!h]
\centering
\includegraphics[width=0.8\textwidth]{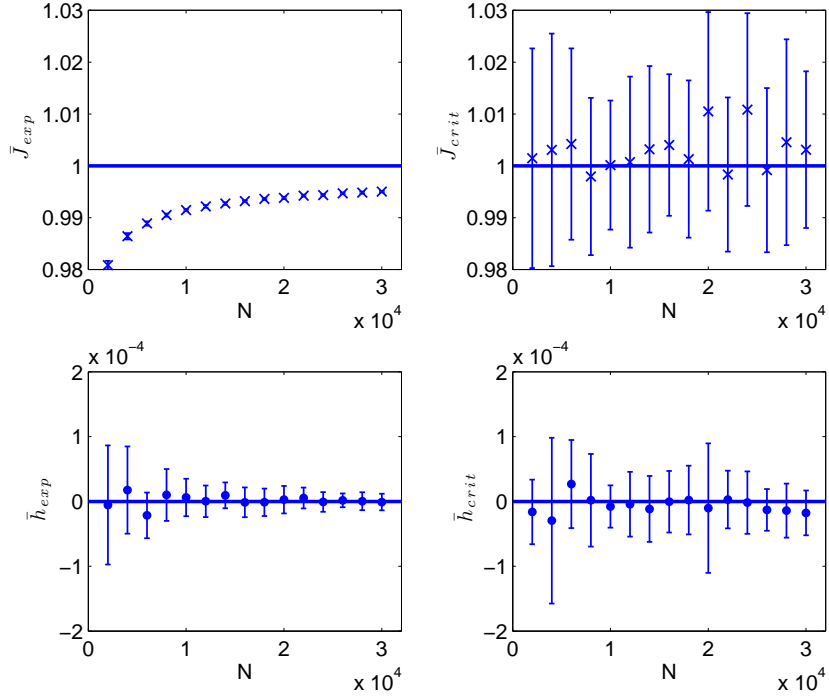}
\caption{\label{Fig18} $J=1$, $h=0$. Left panels: values of $\bar{J}_{exp}$ (upper panel) and $\bar{h}_{exp}$ (lower panel) as a function of $N$ obtained with the standard inversion equations (\ref{stimatori}) and (\ref{stimatorih}).
Rigth panels:  values of $\bar{J}_{crit}$ (upper panel) and $\bar{h}_{crit}$ (lower panel) as a function of $N$ obtained with the Eq. (\ref{Jcrit}) for $J_{crit}$ and (\ref{hcrit}) for $h_{crit}$. 
For all panels: error bars are standard deviations on $20$ different $M$-samples of the same system ($M=1000$ - see text for the details of the simulation); the horizontal lines correspond to the exact values of $J=1$ (upper panels) and $h=0$ (lower panels).}
\end{figure}

\begin{figure}[!h]
\centering
\includegraphics[width=0.8\textwidth]{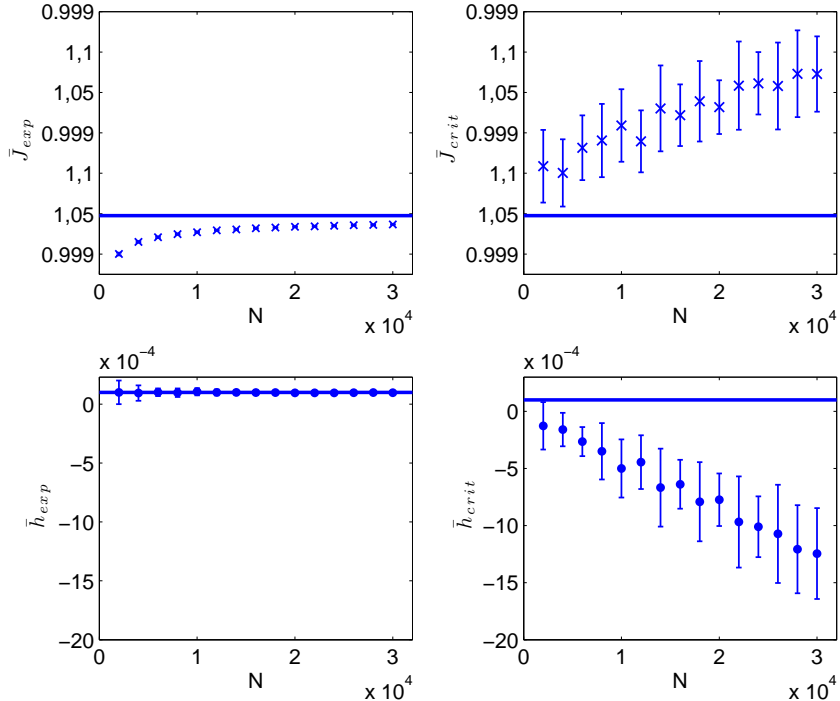}
\caption{\label{Fig19}$J=0.999$, $h=0.0001$. Left panels: values of $\bar{J}_{exp}$ (upper panel) and $\bar{h}_{exp}$ (lower panel) as a function of $N$ obtained with the standard inversion equations (\ref{stimatori}) and (\ref{stimatorih}).
Rigth panels:  values of $\bar{J}_{crit}$ (upper panel) and $\bar{h}_{crit}$ (lower panel) as a function of $N$ obtained with the Eq. (\ref{Jcrit}) for $J_{crit}$ and (\ref{hcrit}) for $h_{crit}$. 
For all panels: error bars are standard deviations on $20$ different $M$-samples of the same system ($M=1000$ - see text for the details of the simulation); the horizontal lines correspond to the exact values of $J=0.999$ (upper panels) and $h=0.0001$ (lower panels).}
\end{figure}

When $J =1$ and $h=0$ the inversion formulas (\ref{stimatori}) and (\ref{stimatorih}) do not hold true because the susceptibility $\chi$ grows to infinity.
Nevertheless, it is still possible to write an expression of the model's parameters in terms of experimental data that do not involve the susceptibility (see the appendix for details). In particular, the expression for the interacting parameter, analogous to Eq.~(\ref{stimatori}), is
\begin{equation}\label{Jcrit}
J_{crit}=\sqrt{\frac{\Gamma^{2}(3/4)\sqrt{6}}{\pi\hat{\chi}_{exp}\sqrt{(1-m^2_{exp})(1-3m^2_{exp})}}} 
\end{equation}
where $\Gamma$ denotes the Gamma function and
\begin{equation}
\hat{\chi}_{exp}=\sqrt{N}\left(\dfrac{1}{M}\sum\limits_{s=1}^{M}m_{N}^{2}(\sigma^{(s)})-m_{exp}^{2}\right),
\end{equation}
while the corresponding of Eq.~(\ref{stimatorih}) for $h_{crit}$ is obtained by inverting Eq.~(\ref{eqcons}) with $J=J_{crit}$:
\begin{equation}\label{hcrit}
h_{crit} =\tanh^{-1}(m_{exp})-J_{crit}m_{exp}.
\end{equation}
In Fig.~\ref{Fig18}, we compare the parameters values obtained using formulas (\ref{Jcrit}) and (\ref{hcrit}) - right panels - with those computed with the inversion formulas (\ref{stimatori}) and (\ref{stimatorih}) - left panels. Note the performance of the expression (\ref{Jcrit}), that predicts the correct value $J=1$ also with a small number of particles, while the standard estimator (\ref{stimatori}) underestimates the exact value of $J$ for all the considered values of $N$.
Despite these good results, it is worth to mention that such critical formulas are not really useful to solve the inverse problem starting from real empirical data because they hold true only in the critical case $J=1$ and $h=0$. This means that when $J\neq 1$ and $h\neq 0$ they fail in reconstructing the parameters values, as shown in Fig.~\ref{Fig19} for the case $J=0.999$ and $h=0.0001$. In fact, while the standard inversion equation underestimates the exact value of $J$ as in the previous example (Fig.~\ref{Fig18}), Eq.~(\ref{Jcrit}) overestimates it with an error that grows as the number of particles $N$ increases. As a consequence of this bad estimation of the couplings, the error in the reconstruction of the magnetic field with Eq.~(\ref{hcrit}) is big too. Therefore, expressions (\ref{Jcrit}) and (\ref{hcrit}) can not be apply with real data outside criticality, but they can be used as a tool to measure if the data come from a system that is really critical or only near to criticality.

\section{Inverse Problem for the Multi-Species Model}
In many real-world studies (e.g. in socio-economic, biological or neuro-physical sciences), there are situations in which the problem is to model a mean field interacting system partitioned into different sets where the elements (individuals, agents or neurons) belonging to the same set share very similar features or attributes.
Formally, such a model can be thought of as an extension of the Curie-Weiss model to systems composed of many interacting groups in the following way\cite{FVC}:
let us consider a system of $N$ particles that can be divided into $k$ subsets $P_{1},\dots , P_{k}$ with $P_{l}\cap P_{s}=\emptyset$, for $l\neq s$ and sizes $|P_{l}|=N_{l}$, where $\sum_{l=1}^{k}N_{l}=N$. Particles interact 
with each other and with an external field according to the mean field Hamiltonian:
\begin{equation}\label{Hami.multi}
H_{N}(\boldsymbol{\sigma})=-\frac{1}{2N}\sum_{i,j=1}^{N}J_{ij}\sigma_{i}\sigma_{j}-\sum_{i=1}^{N}h_{i}\sigma_{i} \; 
\end{equation}
where $\sigma_{i}\in\{+ 1, -1\}$ represents the spin of the particle $i$, $J_{ij}$ is the parameter that tunes the mutual 
interaction between the particle $i$ and the particle $j$ and $h_i$ is the $i$-th external magnetic field. $J_{ij}$ and $h_i$ take values according to the following symmetric matrix and vector, respectively:

\begin{displaymath}
         \begin{array}{l}
                \vspace{0.1cm}\\
                N_1 \left\{ \begin{array}{ll|||}
                                      \\
                                   \end{array}  \right.
                                        \\
                N_2  \left\{ \begin{array}{ll|||}
                                        \\
                                   \end{array}  \right. 
                                          \\
                          \\
                         \vspace{0.15cm} \\
                     
                  N_k \left\{ \begin{array}{ll|||}
                     \\
            \\
                \\
                                  \end{array}  \right.
                               \\
         \end{array}
          \!\!\!\!\!\!
         \begin{array}{ll||}
                \;\;\;\;
                 \overbrace{\;\;\;\;\;\; }^{\textrm{$N_1$}}
                 \overbrace{\;\;\;\;\;\; }^{\textrm{$N_2$}}\quad\quad\quad\quad
                 \overbrace{\quad\quad\quad\quad\;\; }^{\textrm{$N_k$}}
                  \\
                 \left(\begin{array}{c|c|cc|ccc}
                               \mathbf{ J}_{11}  &  \mathbf{ J}_{12} & &\;\dots\; & &\;\;\mathbf{ J}_{1k}\;\;&
                                \\
                                 \hline
                              \mathbf{ J}_{12} & \mathbf{ J}_{22} & & & & &\\
                             \hline
                             & & & & & &\\
                             \vdots & & & & & &\\
                             \hline
                             & & & & & &\\
                             \mathbf{ J}_{1k} & \mathbf{ J}_{2k} & &\;\dots\; & &\;\;\mathbf{ J}_{kk}\;\; &\\
                             & & & & & &
                      \end{array}\right)
               \end{array}
\qquad
         \begin{array}{l}
                \vspace{0.1cm}\\
                N_1 \left\{ \begin{array}{ll|||}
                                      \\
                                   \end{array}  \right.
                                        \\
                N_2  \left\{ \begin{array}{ll|||}
                                        \\
                                   \end{array}  \right. 
                                          \\
                          \\
                         \vspace{0.15cm} \\
                     
                  N_k \left\{ \begin{array}{ll|||}
                     \\
            \\
                \\
                                  \end{array}  \right.
                               \\
         \end{array}
          \!\!\!\!\!\!
\begin{array}{ll||}
       \\
    \left(\begin{array}{ccc|c}
                \mathbf{h}_{1}
            \\
            \hline
            
            \mathbf{h}_{2}
            \\
            \hline
            \\
            \vdots
            \\
            \hline
            \\
            \mathbf{h}_{k}
            \\
            \\
        \end{array}\right)
               \end{array}
\end{displaymath}\\
\noindent where each block $\mathbf{J}_{ls}$ has constant elements $J_{ls}$ and each $\mathbf{h}_{l}$ is a vector of constant elements $h_{l}$. For $l=s$, $\mathbf{J}_{ll}$ is a square matrix, whereas the matrix $\mathbf{ J}_{ls}$ is rectangular. We assume $J_{11}, J_{22},\dots , J_{kk}$ to be positive, whereas $J_{ls}$ with $l\neq s$ can be either positive or negative allowing for both ferromagnetic and antiferromagnetic interactions. The different values of the vector field depend on the subset the particles belong to.

Indicating with $m_{l}(\boldsymbol{\sigma})$ the total magnetization of the group $P_{l}$, and with $\alpha_{l}=N_{l}/N$ the relative size of the set $P_{l}$, we may easily express the Hamiltonian (\ref{Hami.multi}) as:
\begin{align}\label{Hami.multi.2}
H_{N}(\boldsymbol{\sigma}) &=-N\Big(\frac{1}{2}\sum\limits_{l, s=1}^{k}\alpha_{l}\alpha_{s}J_{ls}m_{l}(\boldsymbol{\sigma})m_{s}(\boldsymbol{\sigma})+\sum\limits_{l=1}^{k}\alpha_{l}h_{l}m_{l}(\boldsymbol{\sigma})\Big)\nonumber\\
&=-N\Big(\frac{1}{2}\langle\mathbf{J}\mathbf{D}_{\boldsymbol{\alpha}}\mathbf{m}(\boldsymbol{\sigma}),\mathbf{D}_{\boldsymbol{\alpha}}\mathbf{m}(\boldsymbol{\sigma})\rangle +\langle \mathbf{h}, \mathbf{D}_{\boldsymbol{\alpha}}\mathbf{m}(\boldsymbol{\sigma})\rangle\Big)
\end{align}
where $\mathbf{m}(\boldsymbol{\sigma})=(m_{1}(\boldsymbol{\sigma}),\dots ,m_{k}(\boldsymbol{\sigma}))$, $\mathbf{D}_{\boldsymbol{\alpha}}=diag\{\alpha_{1},\dots,\alpha_{k}\}$, $\mathbf{h}=(h_{1},\dots , h_{k})$ and $\mathbf{J}$ is the reduced interaction matrix
\begin{equation*}
\mathbf{J}=\begin{pmatrix}
J_{11}  & J_{12} & \dots & J_{1k}\\
J_{12}  & J_{22} & \dots & J_{2k}\\
\vdots&\vdots&&\vdots \\
J_{1k}  & J_{2k} & \dots & J_{kk}
\end{pmatrix}.
\end{equation*}
The joint distribution of a spin configuration $\boldsymbol{\sigma}=(\sigma_{1},\dots ,\sigma_{N})$ is given by the Boltzmann-Gibbs measure $P_{N, \mathbf{J}, \mathbf{h}}$ related to the Hamiltonian (\ref{Hami.multi}), where again we consider the inverse temperature parameter $\beta$ absorbed within the model parameters $\mathbf{J}$ and $\mathbf{h}$.  The model is well-posed, as it has been shown in Ref.~\cite{GC}.
In the thermodynamic limit the model is described by the following system of mean-field equations:
\begin{equation}\label{campomedio.multi}
\begin{cases}
m_{1}(\mathbf{J},\mathbf{h}) &\!\!\!\!= \tanh\Big(\sum\limits_{l=1}^{k}\;\alpha_{l}J_{1l}\;m_{l}(\mathbf{J},\mathbf{h})+h_{1}\Big) \\
m_{2}(\mathbf{J},\mathbf{h}) &\!\!\!\!=\tanh\Big(\sum\limits_{l=1}^{k}\;\alpha_{l}J_{2l}\;m_{l}(\mathbf{J},\mathbf{h})+h_{2}\Big)\\
\;\vdots\\
m_{k}(\mathbf{J},\mathbf{h}) &\!\!\!\!=\tanh\Big(\sum\limits_{l=1}^{k}\;\alpha_{l}J_{lk}\;m_{l}(\mathbf{J},\mathbf{h})+h_{k}\Big) \; .
\end{cases} 
\end{equation}
In particular, the solutions of this system are the critical points of the pressure function of the model (see Ref.~\cite{GC}). 
When the system admits a unique thermodynamically stable solution $\mathbf{m}(\mathbf{J},\mathbf{h})=(m_{1}(\mathbf{J},\mathbf{h}),\dots,m_{k}(\mathbf{J},\mathbf{h}))$, the inversion problem procedure is the natural extension of the case we have studied for the Curie Weiss model when the Boltzmann-Gibbs distribution of the total magnetization is unimodal. Following the study of Ref.~\cite{FVC} where this case has been analyzed, we denote by $m_{l\; exp}$ the average magnetization of each specie calculated from the data 
\begin{equation*}
m_{l\; exp}=\dfrac{1}{M}\sum\limits_{i=1}^{M}m_{l}(\boldsymbol{\sigma}^{(i)})\quad l=1,\dots,k
\end{equation*}
and we define the matrices $\mathbf{P}_{exp}=diag\{1-m_{1\; exp}^{2},\dots,1-m_{k\; exp}^{2}\}$ and $\boldsymbol{\chi}_{exp}$, whose elements are 
\begin{equation*}
\chi_{lr\; exp}=N_{r}\left(\dfrac{1}{M}\sum\limits_{i=1}^{M}m_{l}(\boldsymbol{\sigma}^{(i)})m_{r}(\boldsymbol{\sigma}^{(i)})-m_{l\; exp}m_{r\; exp}\right)\quad l,r=1,\dots,k.
\end{equation*}
The model estimators are
\begin{align}\label{StimatoribipJ}
\mathbf{J}_{exp} &=(\mathbf{P}_{exp}^{-1}-\boldsymbol{\chi}_{exp}^{-1})\mathbf{D}_{\boldsymbol{\alpha}}^{-1}\\
h_{l\; exp} &=\tanh^{-1}(m_{l\; exp})-\sum\limits_{r=1}^{k}\;\alpha_{r}J_{lr\; exp}m_{r\; exp}\quad l=1,\dots,k.\label{Stimatoribiph}
\end{align}
The parameter reconstruction for this case has been deeply investigated in Ref.~\cite{FVC}. In the following section we consider  cases in which the system of mean-field equations (\ref{campomedio.multi}) has more stable (or metastable) solutions; in these situations Eq.~(\ref{StimatoribipJ}) and (\ref{Stimatoribiph}) fail to provide a good parameter reconstruction. Nevertheless, since the previous equations
 are locally fulfilled around each solution, the inverse problem can be globally solved by applying the analogous procedures to those described for the Curie-Weiss model as the consistence equation admits more solutions. 
 
Without loss of generality, we will present the results only for the two-species case ($k=2$). This choice is motivated by the fact that a big number of species would cause a loss of statistical robustness working with real world datasets and an excessive increase of computational complexity in the case of numerical simulations.

\subsection{Numerical Tests}

As a test problem for the multi-species mean-field model we consider systems of $N\in [200, 2000]$ particles divided into $k=2$ equally populated subsets ($N_1=N_2=N/2$) and a sample of $M=1000$ independent spin configurations. Starting from couples of given values for the reduced interaction matrix 
and for the external vector field 

\begin{equation}
\mathbf{J}=
\begin{pmatrix}
    J_{11}  &  J_{12} \\
    J_{12}  &  J_{22} 
\end{pmatrix}
\qquad \qquad
\mathbf{h}=
\begin{pmatrix}
    h_1 \\
    h_2
\end{pmatrix}
\end{equation}
\noindent
we consider $20$ $M$-samples for each couple $(\mathbf{J},\mathbf{h})$ and we apply the maximum likelihood estimation to each one of them independently; then we average the inferred values $\mathbf{J}_{exp}$ and $\mathbf{h}_{exp}$ of the model parameters, given by (\ref{StimatoribipJ}) and (\ref{Stimatoribiph}), over the $20$ $M$-samples
(as in the Curie-Weiss model) obtaining $\mathbf{\bar{J}}_{exp}$ and $\mathbf{\bar{h}}_{exp}$.

\subsubsection{Distribution with 2 or more peaks}
Let us consider the case in which the system (\ref{campomedio.multi}) admits three solutions $(m_{1}(\mathbf{J},\mathbf{h}), m_{2}(\mathbf{J},\mathbf{h}))$, corresponding to two maxima and a minimum of the pressure function. Consider as an example:
\begin{equation}
\mathbf{J}=
\begin{pmatrix}
   1.4  &  0.98 \\
    0.98 &  1.4 
\end{pmatrix}
\qquad \qquad
\mathbf{h}=
\begin{pmatrix}
   0.001 \\
   0.002
\end{pmatrix}.
\end{equation}
\begin{figure}[!h]
\centering
\includegraphics[width=0.8\textwidth]{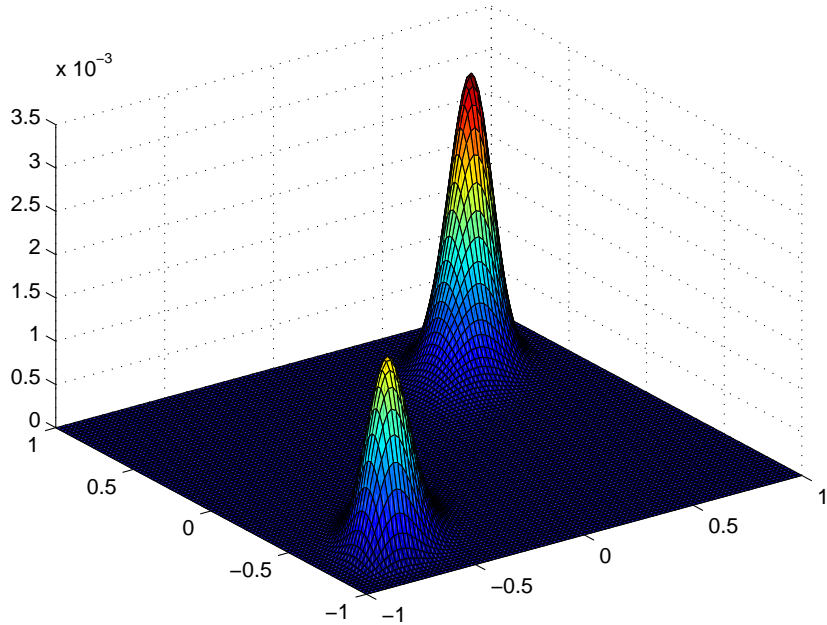}
\caption{\label{Fig21} Boltzmann-Gibbs distribution of the total magnetization for $J_{11}=J_{22}=1.4$, $J_{12}=0.98$, $h_1=0.001$, $h_2=0.002$ and $N_1=N_2=N/2=100$ spins. The system (\ref{campomedio.multi}) admits three solutions $(m_{1}(\mathbf{J},\mathbf{h}), m_{2}(\mathbf{J},\mathbf{h}))$, corresponding to two maxima and a minimum of the pressure function.}
\end{figure}

\noindent
In this case the Boltzmann-Gibbs distribution of the total magnetization presents two peaks, one in correspondence to the local maximum $(m_{1}(\mathbf{J},\mathbf{h})=-0.6436, m_{2}(\mathbf{J},\mathbf{h})=-0.6432)$ and one in correspondence to the global maximum $(m_{1}(\mathbf{J},\mathbf{h})=0.6489, m_{2}(\mathbf{J},\mathbf{h})=0.6496)$, as shown in Fig.~\ref{Fig21}. 

\begin{figure}[!h]
\centering
\includegraphics[width=0.8\textwidth]{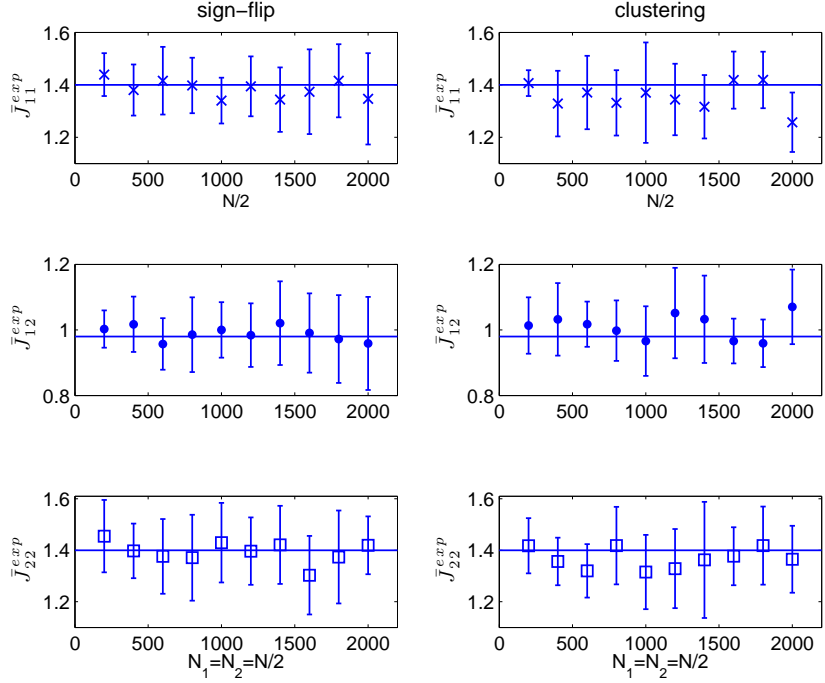}
\caption{\label{Fig21a} Elements of the matrix $\mathbf{\bar{J}}_{exp}$ as a function of $N$ for $J_{11}=J_{22}=1.4$, $J_{12}=0.98$, $h_1=0.001$ and $h_2=0.002$. The values of  $\bar{J}_{11}^{exp}$ (crosses), $\bar{J}_{12}^{exp}$ (dots) and $\bar{J}_{22}^{exp}$  (squares) in the left panels are obtained with the sign-flip, those of right panels with the clustering algorithm. The horizontal lines correspond to the exact values of the elements of the matrix $\mathbf{J}$.}
\end{figure}

\begin{figure}[!h]
\centering
\includegraphics[width=0.8\textwidth]{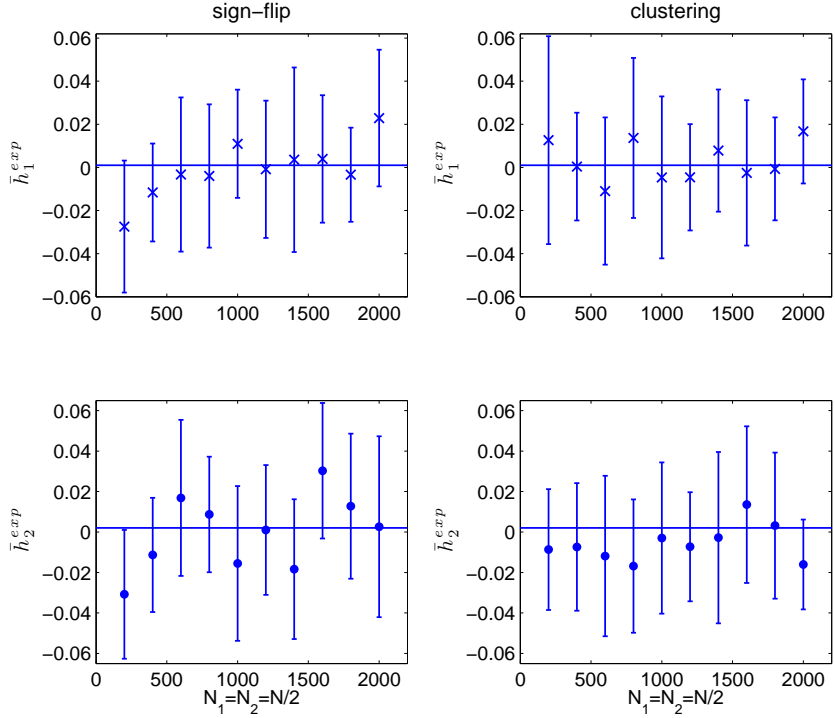}
\caption{\label{Fig21b} Elements of $\mathbf{\bar{h}}_{exp}$ as a function of $N$ for $J_{11}=J_{22}=1.4$, $J_{12}=0.98$, $h_1=0.001$ and $h_2=0.002$. The values of $\bar{h}_1^{exp}$ (crosses) and $\bar{h}_2^{exp}$ (dots) in the left panels are obtained with the sign-flip, those of right panels with the clustering algorithm. The horizontal lines correspond to the exact values of the elements of $\mathbf{h}$.}
\end{figure}

Figs.~\ref{Fig21a} and \ref{Fig21b} represent the reconstruction of the model parameters using both the sign-flip trick (left panels) and clustering algorithm (right panels). 

The results obtained in both cases fully satisfy the expectation also for groups with few elements ($N<1000$). The advantage of the sign-flip with respect to the clustering algorithm is of computational type.

\begin{figure}[!h]
\centering
\includegraphics[width=0.8\textwidth]{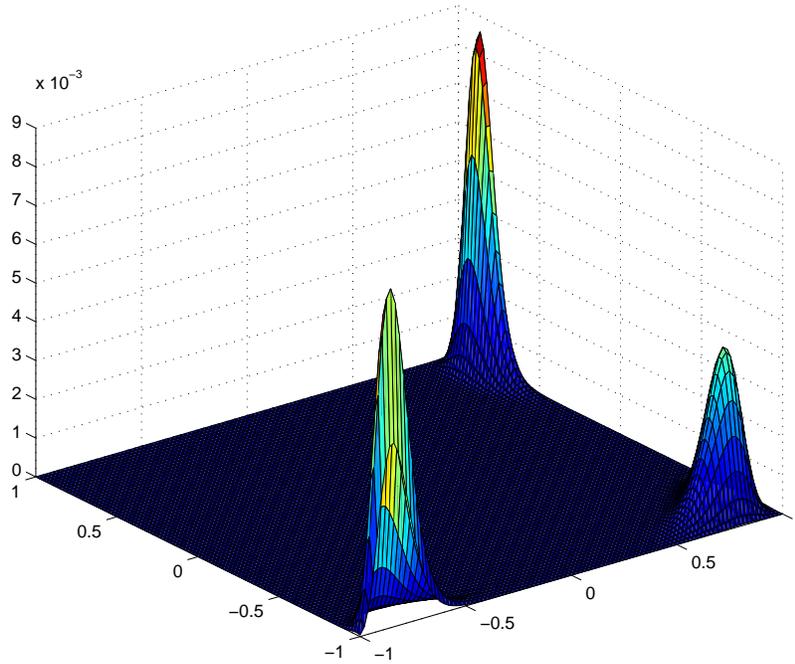}
\caption{\label{Fig22} Boltzmann-Gibbs distribution of the total magnetization for $J_{11}=J_{22}=2.8$, $J_{12}=0.7$ and $h_1=-h_2=0.32$ and $N_1=N_2=N/2=100$ spins. The system (\ref{campomedio.multi}) admits five solutions $(m_{1}(\mathbf{J},\mathbf{h}), m_{2}(\mathbf{J},\mathbf{h}))$, three of wich are  maxima of the pressure function.}
\end{figure}

The clustering algorithm becomes essential when the maxima of the pressure function are more than two because in these cases the Boltzmann-Gibbs distribution of the total magnetization can not be reduced to a unimodal one through a simple change of sign. Fig.~\ref{Fig22} is an example of this situation for 
\begin{equation}
\mathbf{J}=
\begin{pmatrix}
   2.8  &  0.7 \\
    0.7 &  2.8 
\end{pmatrix}
\qquad \qquad
\mathbf{h}=
\begin{pmatrix}
   0.32 \\
   -0.32
\end{pmatrix}.
\end{equation}

\begin{figure}[!h]
\centering
\includegraphics[width=0.8\textwidth]{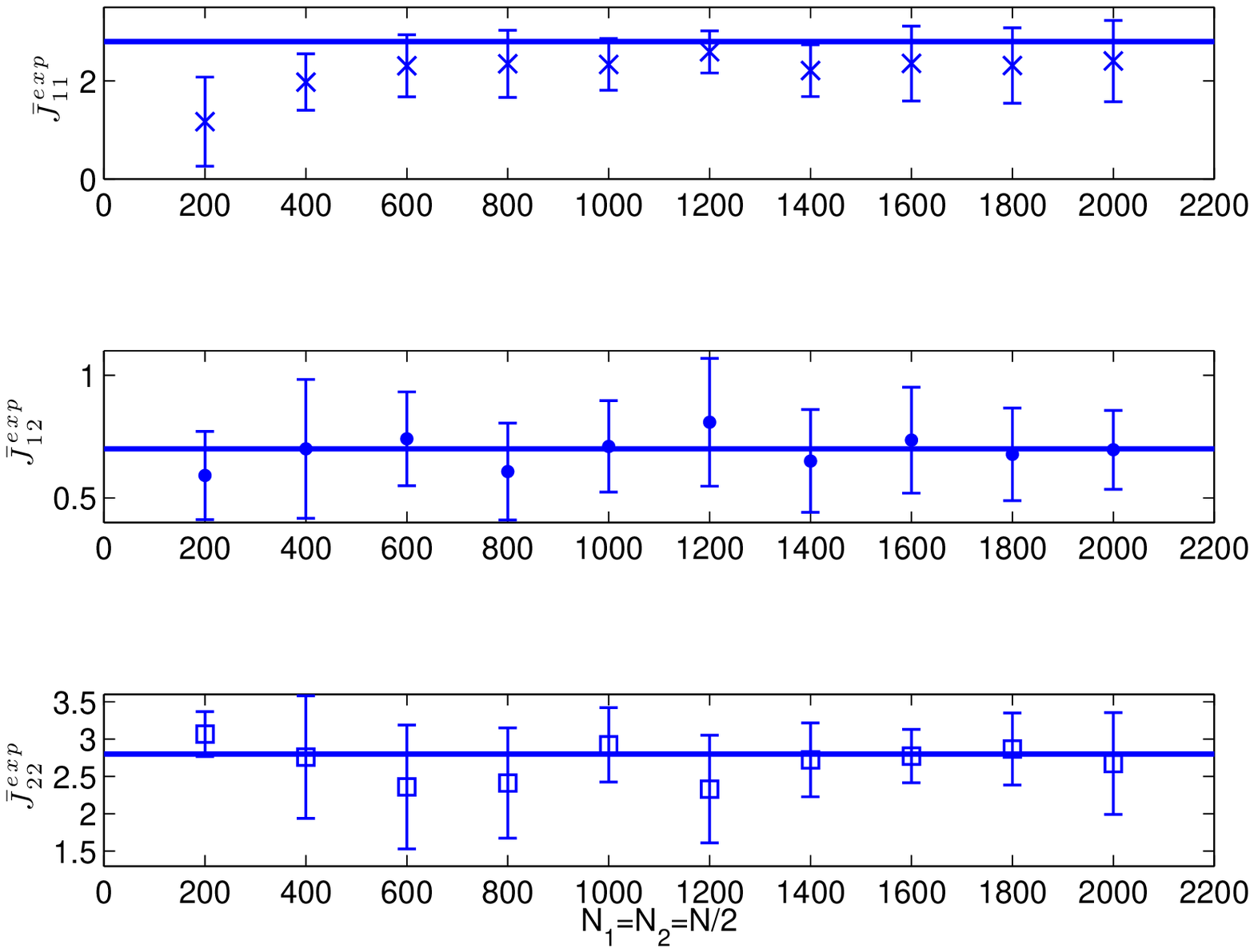}
\caption{\label{Fig22a} Elements of the matrix $\mathbf{\bar{J}}_{exp}$ as a function of $N$ for $J_{11}=J_{22}=2.8$, $J_{12}=0.7$ and $h_1=-h_2=0.32$ obtained with clustering algorithm. The horizontal lines correspond to the exact values of the elements of the matrix $\mathbf{J}$.}
\end{figure}

\begin{figure}[!h]
\centering
\includegraphics[width=0.8\textwidth]{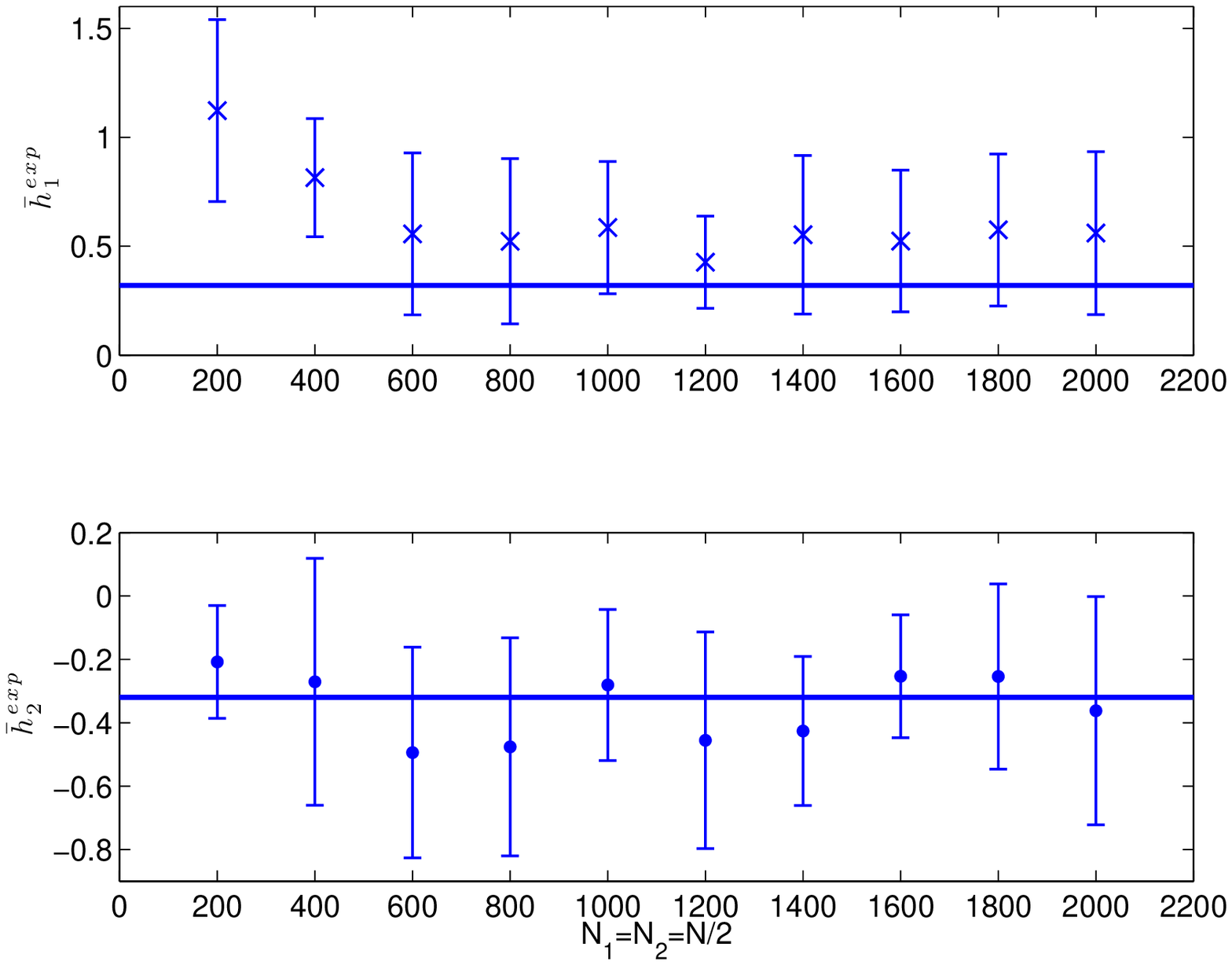}
\caption{\label{Fig22b} Elements of $\mathbf{\bar{h}}_{exp}$ as a function of $N$ for $J_{11}=J_{22}=2.8$, $J_{12}=0.7$ and $h_1=-h_2=0.1$ obtained with clustering algorithm. The horizontal lines correspond to the exact values of the elements of $\mathbf{h}$.}
\end{figure}

\noindent
The parameter reconstruction with the clustering algorithm is shown in Fig.~\ref{Fig22a} and \ref{Fig22b}. The figures show that $N$ greater the $400$ suffices to obtain a good parameter estimation.
 
\section{Conclusions}
In this paper we studied the inverse problem for the Curie-Weiss model and for its multi-species version in the low temperature phase, where more than one state is present. In order to infer the parameters of the underlying model starting from input data with two or more coexisting states, we used the well known clustering algorithm and/or the sign-flip of the experimental magnetizations. The predictions of the model parameters produced in these two ways are comparable and very accurate even when the size of the system is small, but when the symmetry of the states in the phase space allows the application, the sign-flip is preferable because is simpler and has a lower computational cost.
Given a set of input configurations with magnetizations either positive and negative, it is necessary before applying the inversion procedure to change the sign of the magnetizations from positive to negative (or viceversa), in order to have in the input only concordant magnetizations.
This work shows results that are particularly useful in applications to real world dataset. It explains, for example, that the sign of the reconstructed magnetic field, contrarily to a common expectation, could not be in accordance with that of the sampled magnetization. This happens when the distribution of the magnetization of the underlying model is multimodal and the input configurations come from the set with smaller probability. Moreover, the expressions of the parameters given for the Curie-Weiss model at the criticality are useful for determining whether a system is in a critical regime or not.

\appendix

\section{Appendices}
Here we describe how to obtain the critical expressions (\ref{Jcrit}) and (\ref{hcrit}) shown in section \ref{CWcriticoSec}. To this purpose it is worth to mention that the reconstruction of the model parameters from data is based on the possibility to find a suitable normalization of the total magnetization $m_{N}(\sigma)$ that remains a well defined random variable also in the thermodynamic limit. Outside of the critical point, the answer of this problem is given by the random variable
\begin{equation}
X=N^{1/2} m_N(\sigma)
\end{equation}
whose distribution in the thermodynamic limit is a Gaussian with mean equal to the stable solution $m$ of the consistence equation \ref{eqcons} and variance equal to the susceptibility $\chi$ of the model\cite{EN,ENR}. Since 
\begin{equation}
Var(X)=NVar(m_N(\sigma)) \rightarrow \chi= \dfrac{1-m^{2}}{1-J(1-m^{2})} \quad \text{as } N\rightarrow\infty,
\end{equation}
by inverting this limit identity and remembering that $m$ is also the limit value of $\langle m_N(\sigma)\rangle$,  we get the inversion formula for the interaction parameter:
\begin{equation}
J=\dfrac{1}{1-\langle m_N^{2}(\sigma)\rangle} - \dfrac{1}{NVar(m_N(\sigma))}.
\end{equation}

When $J=1$ and $h=0$, $X$ is no more a well define random variable in the limit because $\chi$ grows to infinity. In this case the correct normalization of $m_{N}(\sigma)$ is given by
\begin{equation}
\hat{X}=N^{1/4} m_N(\sigma)
\end{equation}
distributed in the thermodynamic limit as follows:
\begin{equation}
\frac{\exp(\frac{1}{4!}\frac{\partial^4p}{\partial x^4}(m)x^4)dx}{\int \exp(\frac{1}{4!}\frac{\partial^4p}{\partial x^4}(m)x^4)dx}
\end{equation}
where $m$ is the unique stable solution of the consistence equation and
\begin{equation}
p(x)=-\frac{J}{2}x^{2}+\ln(\cosh(Jx+h))
\end{equation}
is the pressure function of the model\cite{EN,ENR}.
It is straightforward to show that $m$ is the global maximum point of $p$ and is equal to zero. 
In the limit the variance of $\hat{X}$ is 
\begin{equation}
\lim_{N\rightarrow\infty}Var(\hat{X})=\hat{\chi}=\sqrt{\frac{4!}{-\frac{\partial^4p}{\partial x^4}(0)}}\frac{\Gamma^{2}(3/4)}{\pi\sqrt{2}}
\end{equation}
where 
\begin{equation}
\frac{\partial^4p}{\partial x^4}(x)=-2J^4(1-\tanh^2(Jx+h))(1-3\tanh^2(Jx+h)).
\end{equation}
This means that as $N\rightarrow\infty$ the following identity holds true
\begin{equation}
\frac{\Gamma^{2}(3/4)}{\pi\sqrt{2}} \sqrt{\frac{4!}{2J^4(1-m^2)(1-3m^2)}}=\sqrt{N}Var(m_N(\sigma))
\end{equation}
It follows:
\begin{equation}
J=\sqrt{\frac{\Gamma^{2}(3/4)\sqrt{6}}{\pi Var(m_N(\sigma))\sqrt{N(1-\langle m_N^{2}(\sigma)\rangle)(1-3\langle m_N^{2}(\sigma)\rangle)}}}. 
\end{equation}

\section*{Acknowledgment}
The authors thank P. Contucci for inspiring this work and
 C. Giberti for interesting discussions and  for a careful reading of the manuscript.
M. Fedele thanks the INdAM-COFUND Marie Curie fellowships for financial support.
This work was partially
supported by
FIRB Grant RBFR10N90W.

\end{document}